# On The Multi-Hop Extension of Energy-Efficient WSN Time Synchronization Based on Time-Translating Gateways

Qimeng Liao and Kyeong Soo Kim, *Member*, *IEEE*
Department of Electrical and Electronic Engineering, Xi'an Jiaotong-Liverpool University
Suzhou, 215123, P. R. China
Email: Qimeng.Liao14@student.xjtlu.edu.cn, Kyeongsoo.Kim@xjtlu.edu.cn

*Abstract*— We report preliminary results of a simulation study on the multi-hop extension of the recently-proposed energy-efficient wireless sensor network time synchronization scheme based on time-translating gateways. Unlike the single-hop case, in multi-hop time synchronization a sensor node sends measurement data to a head node through gateways which translate the timestamp values of the received measurement data. Through simulations for two-hop time synchronization, we analyze the impact of the addition of a gateway and its translation of timestamp values on the clock frequency and measurement time estimation and thereby demonstrate the feasibility of the multi-hop extension of the energy-efficient WSN time synchronization based on time-translating gateways.

Keywords- Time synchronization; source clock frequency recovery; energy efficiency; wireless sensor networks

## I. INTRODUCTION

Wireless sensor networks (WSNs) have been the focus of extensive studies due to their versatility and broad range of applications in many areas including industry and military. Time synchronization is one of critical components in WSN operation, as it provides a common time frame among different nodes. It supports functions such as fusing data from different sensor nodes, time-based channel sharing and media access control (MAC) protocols, and coordinated sleep wake-up node scheduling mechanisms [1].

Because sensor nodes are low-complexity and battery-powered devices, their energy efficiency is a key factor in designing WSN algorithms and protocols. Several energy-efficient time synchronization schemes have been developed for WSNs, two notable examples of which are recursive time synchronization protocol (RTSP) [2] and pairwise broadcast synchronization (PBS) [3]; these schemes reduce the power consumption of all network nodes without differentiating between a head node and sensor nodes in the protocols.

In a typical WSN, however, a master/head node is equipped with a powerful processor, connected to both wired and wireless networks, and supplied power from outlet because it serves as a gateway between the WSN and a backbone and a center for fusion of sensory data from sensor nodes, which are limited in processing and electrical power because they are connected only through wireless channels and battery-powered.

Note that the recently-proposed energy-efficient WSN time synchronization scheme [4], [5] puts its major focus on this asymmetry, where it concentrates on battery-powered sensor nodes in minimizing energy consumption for time synchronization. The proposed time synchronization scheme is based on the asynchronous source clock frequency recovery (SCFR) [6] and the two-way message exchanges [7] carried out in a reverse way.

In this paper we consider the multi-hop extension of the proposed WSN time synchronization scheme to address the scalability of the original proposal, where a time-translating gateway node is introduced between a head node and sensor nodes, and the details of measurement time estimation in a two-hop WSN is provided with preliminary simulation results for demonstrating its feasibility.

## II. REVIEW OF ENERGY-EFFICIENT WSN TIME SYNCHRONIZATION BASED ON ASYNCHRONOUS SCFR AND REVERSE TWO-WAY MESSAGE EXCHANGES [4]

The major idea of this time synchronization scheme is to allow independent, unsynchronized slave clocks at sensor nodes but running at the same frequency as the reference clock at a head node through the asynchronous SCFR described in [6], which needs only the reception of messages with timestamps at sensor nodes. The clock offset, on the other hand, is estimated at the head node based on the reverse two-way message exchanges as shown in Fig. 1. Compared to the conventional two-way message exchanges shown in Fig. 1 (a), the proposed scheme shown in Fig. 1 (b) does not have periodic, dedicated two-way message exchanges; instead, the "Request" and "Response" synchronization messages are embedded in the most recent timestamped message from the head node and a measurement data report message from a sensor node, respectively. Also, in the reverse two-way message exchanges, it is the head node that sends the "Request" messages, not the sensor node; as a result, the head node knows the current status of the sensor node clock, but the sensor node does not. So the information of sensor node clocks (i.e., clock offsets with respect to the reference clock at the head node) is centrally managed at the head node.

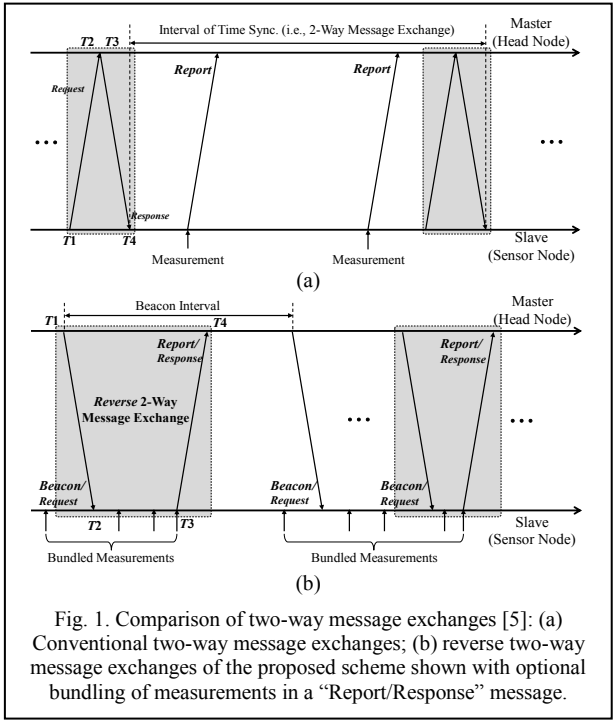

Fig. 1. Comparison of two-way message exchanges [5]: (a) Conventional two-way message exchanges; (b) reverse two-way message exchanges of the proposed scheme shown with optional bundling of measurements in a "Report/Response" message.

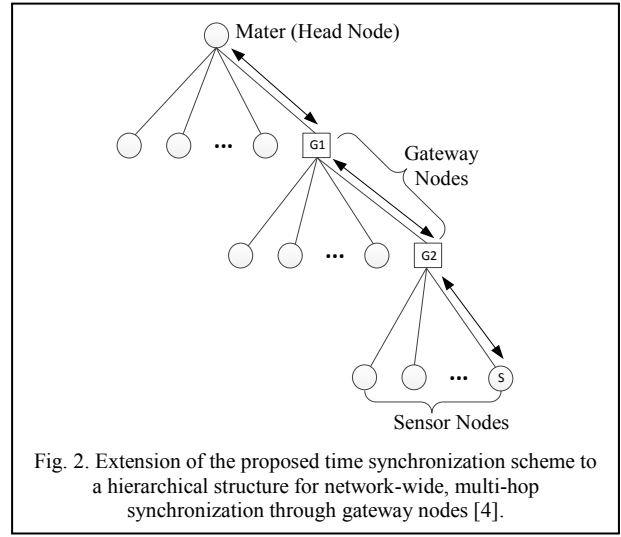

Fig. 2. Extension of the proposed time synchronization scheme to a hierarchical structure for network-wide, multi-hop synchronization through gateway nodes [4].

For operations like coordinated sleep wake-up node scheduling, before sending a control message, the head node adjusts the time for future operation based on the clock offset of the recipient sensor node. In this way, even though sensor nodes in the network have clocks with different clock offsets, their operations can be coordinated based on the common reference clock at the head node.

### III. Two-Hop Time Synchronization

Fig. 2 sketches how the time synchronization scheme described in Section II can be extended to multi-hop synchronization through gateway nodes, which act as both head nodes (for the nodes in the lower hierarchy) and normal sensor nodes (for the gateway node in the higher hierarchy) [4]. Based on this idea, we detail the procedure of time synchronization for a two-hop case. Fig. 3 shows a detailed timing diagram, where a gateway node relays messages and translates their timestamp values between a head node and a sensor node.

Let $t_H$ be the beacon departure time at the head node. Then, the beacon arrival time at the gateway node is expressed as

$$r_{GW\_B\_A} = (t_{H\_B\_D} + D_{HG}) \times R_{HG} + \theta_{HG} + d_{HG},$$

where $R_{HG}$ is the ratio of the gateway node clock frequency to the head node clock frequency, $\theta_{HG}$ is the gateway node clock offset with respect to the head node clock, and $D_{HG}$ and $d_{HG}$ denote a fixed and a noise component of the one-way delay between the head and the gateway node, respectively.

Because the getaway node acts as a head node to the sensor node, the gateway node also sends beacons to the sensor node. Let $t_{GW\_B\_D}$ be the beacon departure time at the gateway node. Note that $t_{GW\_B\_D}$ is based on the gateway node clock. $t_{GW\_B\_D}$ with respect to the head node clock is given by

$$t_{GW\_B\_D}(head) = (t_{GW\_B\_D} - \theta_{HG})/R_{HG}.$$

Then, the beacon arrival time $r_{S\_B\_A}(head)$ at the sensor node with respect to the head node clock is given by

$$r_{S\_B\_A}(head) = t_{GW\_B\_D}(head) + D_{GS} + d_{GS},$$

where $D_{GS}$ and $d_{GS}$ denote a fixed and a noise component of the one-way delay between the gateway and the sensor node, respectively.

Finally, according to the relationship between the sensor node clock and the head node clock, the beacon arrival time $r_{S\_B\_A}$ at the sensor node with respect to its own clock can be formulated as

$$r_{S\_B\_A} = r_{S\_B\_A}(head) \times R_{HS} + \theta_{HS},$$

where $R_{HS}$ is the ratio of the sensor node clock frequency to the head node clock frequency and $\theta_{HS}$ is the sensor clock offset to the head node clock. In the two-hop time synchronization, asynchronous SCFR is being carried out in two different places, i.e., one at the gateway node for the sensor node clock frequency and the other at the head node for the gateway node clock frequency. The cumulative ratio estimator proposed in [6] can be used to estimate $\hat{R}_{GS}$ and $\hat{R}_{HG}$, where $\hat{R}_{GS}$ is the estimated clock frequency ratio between the gateway node and the sensor node, and $\hat{R}_{HG}$ between the gateway node and the head node.

When measurement data are sent from the sensor node via a report packet, its measurement/departure time with respect to the head node clock is denoted by $t_m$, which is to be estimated at the head node, and the departure timestamp $t_{M\_S\_D}$ at the sensor node with respect to its own clock on the condition of the asynchronous SCFR is given by

$$t_{M\_S\_D} = \frac{t_m \times R_{HS} + \theta_{HS} - r_{S\_B\_A}}{\hat{R}_{GS} + r_{S\_B\_A}}.$$

And the measurement data arrival time $t_{M\_GW\_A}$ at the gateway node is formulated as

$$t_{M\_GW\_A} = (t_m + D_{GS}) \times R_{HG} + \theta_{HG} + d_{HG}.$$

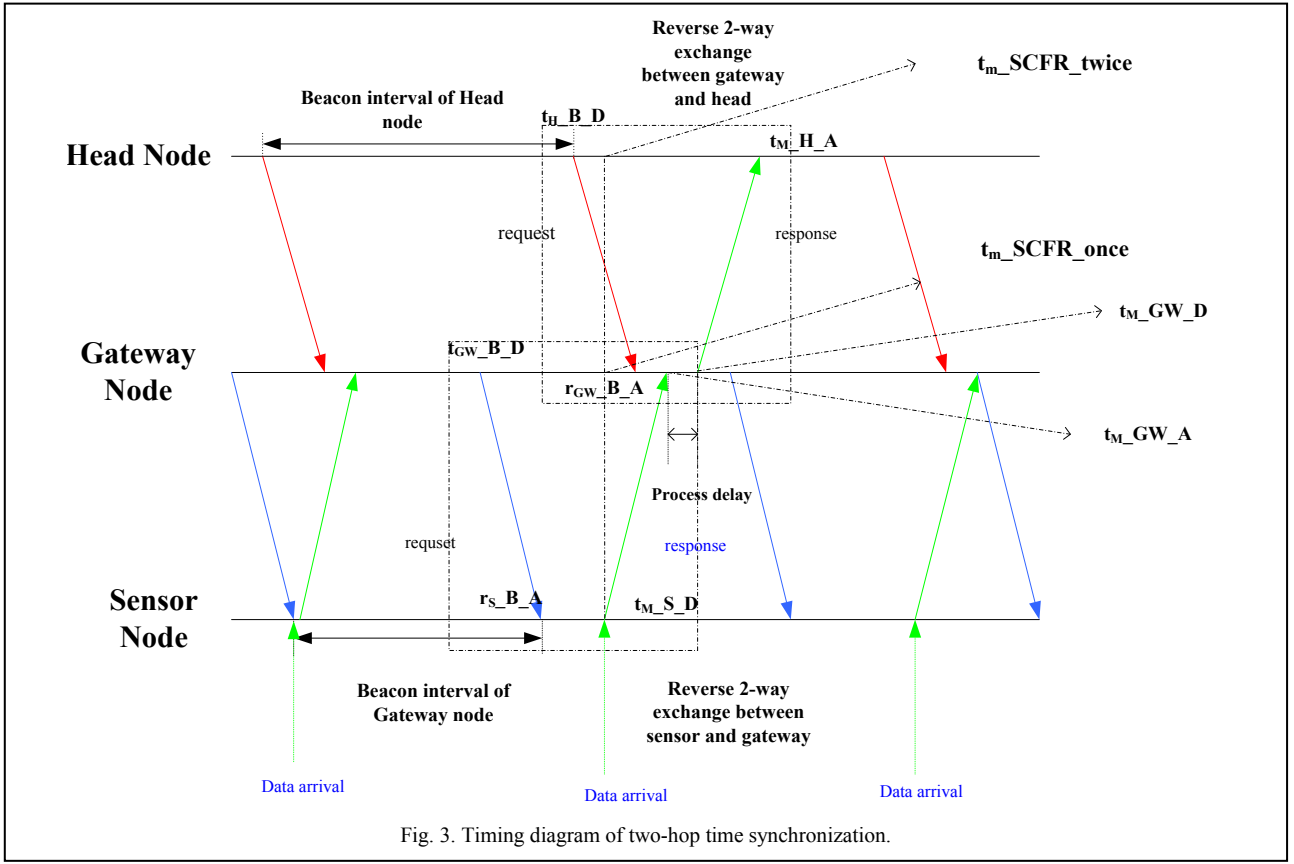

Fig. 3. Timing diagram of two-hop time synchronization.

Therefore, the following formulas can be used to correct the clock offset between the sensor node clock and the gateway node clock:

$$\theta_{est\_}GS = \frac{r_{S\_B\_A} - t_{GW\_B\_D}}{2} - \frac{t_{M\_GW\_A} - t_{M\_S\_D}}{2},$$

$$D_{est\_}GS = \frac{r_{S\_B\_A} - t_{GW\_B\_D}}{2} + \frac{t_{M\_GW\_A} - t_{M\_S\_D}}{2}.$$

Next, the gateway node needs to relay the measurement data to the head node. On the condition of asynchronous SCFR, the received measurement time with respect to the gateway node clock is given by

$$t_M\_SCFR\_once = t_{M\_S\_D} - \theta_{est\_}GS.$$

Note that the head node sends beacons to the gateway node, whose departure timestamp at the head node is $t_{H\_B\_D}$, and arrival timestamp at the gateway node is $r_{GW\_B\_A}$. The departure timestamp of measurement data at the gateway node on the condition of asynchronous SCFR is formulated as:

$$t_M\_GW\_D = \frac{t_{M\_GW\_A} + a - r_{GW\_B\_A}}{\hat{R}_{HG} + r_{GW\_B\_A}},$$

where $a$ means the processing delay of relaying measurement data at the gateway node. The timestamp of measurement data arrival time at the head node is:

$$t_M\_H\_A = \frac{t_{M\_GW\_A} + a - \theta_{HG}}{R_{HG} + D_{HG} + d_{HG}}.$$

Based on the reverse two-way message exchange procedures, we can estimate the clock offset and the propagation delay between the gateway node and the head node as follows:

$$\theta_{est\_}HG = \frac{r_{GW\_B\_A} - t_{H\_B\_D}}{2} - \frac{t_{M\_H\_A} - t_{M\_GW\_D}}{2},$$

$$D_{est\_}HG = \frac{r_{GW\_B\_A} - t_{H\_B\_D}}{2} + \frac{t_{M\_H\_A} - t_{M\_GW\_D}}{2}.$$

On the condition of asynchronous SCFR, therefore, the data measurement time of the received packet with respect to the head node clock is given by

$$t_M\_SCFR\_twice = \frac{t_M\_SCFR\_once - r_{GW\_B\_A}}{\hat{R}_{HG}} + r_{GW\_B\_A} - \theta_{est\_}HG.$$

IV. SIMULATION RESULTS

Here we consider a simple WSN consisting of a sensor node, a gateway node, and a head node. The distance between the sensor node and the gateway node is set to 200 m, and the distance between the head node and the gateway node is set to 100 m. During 120-s observation period, there are totally 100 measurement data—whose arrivals are modeled as a Poisson process—reported from the sensor node to the head node though the gateway node. Clock skew and offset between the sensor node and the head node are 1.0+200 ppm and 0.9 s, respectively. Likewise, the clock skew and offset between the gateway node and the head node are set to 1.0+100 ppm and 1 s, respectively. As for asynchronous SCFR, we use the cumulative ratio estimator [6].

The estimation errors of source clock frequency (i.e., denoted as "frequency difference") and measurement time (i.e., denoted as "measurement time difference") of two-hop time synchronization for the beacon interval of

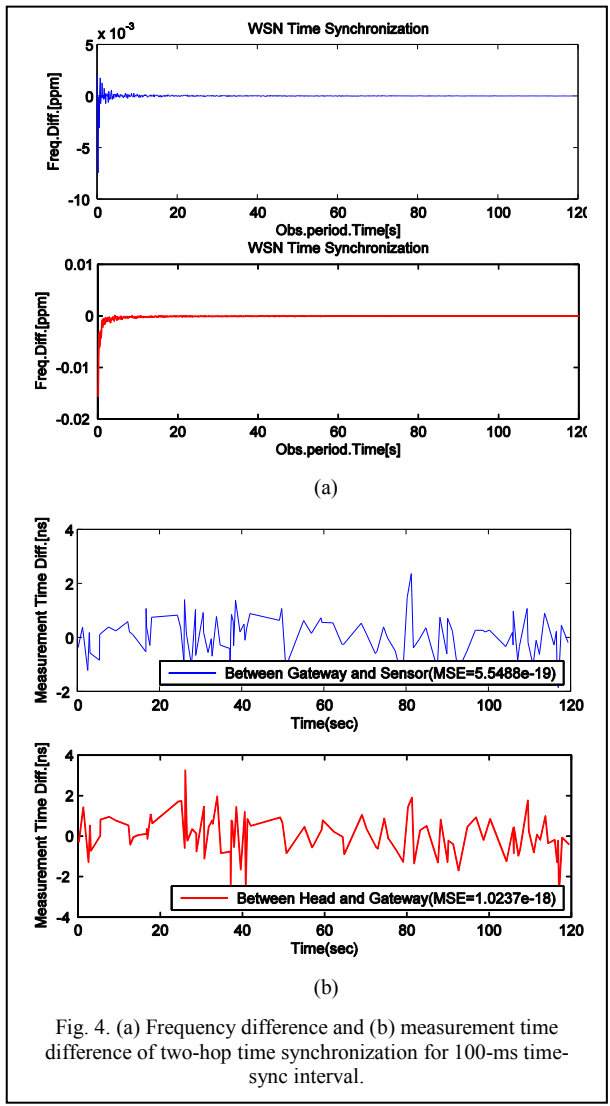

Fig. 4. (a) Frequency difference and (b) measurement time difference of two-hop time synchronization for 100-ms time-sync interval.

100 ms is shown in Fig. 4.[1] We observe that the frequency difference and measurement time difference between the sensor node and the gateway node (i.e., the first plots) are very similar to the simulation results in [8] (i.e. a single-hop case). The results for the head node and the gateway node (i.e., the second plots) also show similar trends. The mean square error (MSE) of measurement time difference for both cases suggests that the cumulative error introduced by a gateway node is minimal, which demonstrates the feasibility of the multi-hop extension of the energy-efficient WSN time synchronization scheme based on asynchronous SCFR and reverse two-way message exchanges.

## V. CONCLUSIONS

We have considered the multi-hop extension of the energy-efficient time synchronization scheme for WSNs based on asynchronous SCFR and reverse two-way message exchanges proposed in [4] and presented the analysis of measurement time estimation in case of two hops. Simulation results for the comparison between single-hop and two-hop time synchronization performance show that the impact of the addition of a gateway and its translation of timestamp values on the clock frequency and measurement time estimation is kept minimal, which demonstrates the feasibility of the multi-hop energy-efficient WSN time synchronization.

ACKNOWLEDGMENT

This work was supported in part by the Centre for Smart Grid and Information Convergence (CeSGIC) at Xi'an Jiaotong-Liverpool University and Xi'an Jiaotong-Liverpool University Research Development Fund under grant reference number RDF-14-01-25.

---

[1] Similar results are observed for different values of beacon intervals, which are not shown here due to space limitations.